\documentclass[letter]{aa}
\usepackage{graphicx}
\usepackage{amssymb}
\usepackage{amsmath}
\usepackage{natbib}
\usepackage{mathrsfs}
\usepackage{ulem}
\usepackage{txfonts}
\usepackage{lmodern}
\usepackage{booktabs}
\usepackage{multicol}
\usepackage{cases}
\usepackage{color}
\usepackage{esint} % various fancy integral symbols
\usepackage{float}

\usepackage[switch]{lineno} % switch=Put the line into the inner margin
\newcommand{\mesa}{{\textsc{MESA}}}
\newcommand{\kepler}{\textit{Kepler}}

\def\m2s2{\,m$^{2}$\,s$^{-2}$} %m2.s -2
\newcommand{\vaisala}{Brunt-V\"ais\"al\"a}

\newcommand{\numax}{\nu_{\rm max}}

\newcommand{\ri}{r_{\rm i}}
\newcommand{\ro}{r_{\rm o}}

\newcommand{\dpun}{\Delta\Pi_1}
\newcommand{\dn}{\Delta\nu}

\newcommand{\domg}{\delta\omega_{\rm g}}

\newcommand{\pg}{P_{\rm g}}
\newcommand{\epsg}{\varepsilon_{\rm g}}
\newcommand{\nmin}{n_{\rm min}}

\newenvironment{itemize*}%
  {\begin{itemize}%
    \setlength{\itemsep}{1pt}%
    \setlength{\parskip}{1pt}}%
  {\end{itemize}}

\newcommand\T{\rule{0pt}{2.6ex}}
\newcommand\B{\rule[-1.2ex]{0pt}{0pt}}

\begin{document}
\title{Strong magnetic fields detected in the cores of 11 red giant stars using gravity-mode period spacings}
\titlerunning{Strong magnetic fields detected in red giant cores}
\author{
S. Deheuvels\inst{1}, G. Li\inst{1}, J. Ballot\inst{1}, F. Ligni\`eres\inst{1}}

\institute{IRAP, Universit\'e de Toulouse, CNRS, CNES, UPS, 31400 Toulouse, France
}

\offprints{S. Deheuvels\\ \email{sebastien.deheuvels@irap.omp.eu}
}

%\date{Submitted ...}

\abstract{Despite their importance in stellar evolution, little is known about magnetic fields in the interior of stars. The recent seismic detection of magnetic fields in the core of several red giant stars has given measurements of their strength and information on their topology. We revisit the puzzling case of hydrogen-shell burning giants that show deviations from the expected regular period spacing of gravity modes. These stars also tend to have a too low measured period spacing compared to their counterparts. We here show that these two features are well accounted for by strong magnetic fields in the cores of these stars. For 11 \kepler\ red giants showing these anomalies, we place lower limits on the core field strengths ranging from 40 to 610 kG. For one star, the measured field exceeds the critical field above which gravity waves no longer propagate in the core. We find that this star shows mixed mode suppression at low frequency, which further suggests that this phenomenon might be related to strong core magnetic fields.}


\maketitle

\section{Introduction \label{sect_intro}}

Magnetic fields affect stars at all evolutionary stages 
from star-forming molecular clouds to white dwarfs and magnetars (\citealt{mckee07}, \citealt{kaspi17}, \citealt{ferrario20}). 
In particular, they are expected to play a central role in the redistribution of angular momentum inside stars (\citealt{maeder05}, \citealt{cantiello14}, \citealt{rudiger15}), and thus in the transport of chemical elements. While surface magnetic fields have been detected and characterized in stars across the HR diagram (\citealt{landstreet92}, \citealt{donati09}), internal magnetic fields have long remained inaccessible to direct observations.
In red giant stars, the detection of mixed modes -- that is, oscillation modes that behave as gravity (g) modes in the core and as pressure modes in the envelope -- has given strong evidence that the cores of red giant stars are rotating slowly (e.g., \citealt{deheuvels12}, \citealt{mosser12b}, \citealt{gehan18}). This yielded evidence that angular momentum is redistributed much more efficiently than if only purely hydrodynamical processes were at work (e.g., \citealt{marques13}). Magnetic fields could produce the additional transport that is needed (\citealt{rudiger15}, \citealt{jouve15}, \citealt{fuller19}, \citealt{petitdemange22}).
Observational constraints on the properties of internal magnetic fields are crucially needed to assess the nature and the efficiency of the magnetic transport of angular momentum inside stars.

The propagation of magneto-gravity waves is expected to be suppressed when the magnetic field exceeds a critical strength $B_{\rm c}$ above which Alfv\'en wave frequencies become comparable to those of gravity waves. This phenomenon was invoked by \cite{fuller15} to account for the unexpectedly low amplitudes of dipole mixed modes in a fraction of red giants. For core fields above $B_{\rm c}$, the authors suggested that the mode energy reaching the magnetized core would be entirely dissipated and lost, giving rise to purely p-like dipole modes. This interpretation was questioned by \cite{mosser17}, who found that partially suppressed dipole modes still retain a g-like character. \cite{loi20a} later showed that even with strong fields, a fraction of the incoming waves could remain g-like, which would allow for partial energy return from the core. The interpretation 
of suppressed dipole modes remains debated.

Magnetic fields also produce shifts in the oscillation mode frequencies (\citealt{gough90}). Several studies have recently investigated the impact of internal fields on the frequencies of mixed modes in red giants (\citealt{gomes20}, \citealt{bugnet21}, \citealt{loi21}). Very recently, \cite{li22} detected clear asymmetries in the rotational multiplets of dipole mixed modes in three \kepler\ red giants. They showed that these features can only be accounted for by internal magnetic fields with intensities ranging from 30 to 130 kG in the vicinity of the hydrogen burning shell. These findings opened the exciting opportunity to characterize magnetic fields in the cores of red giants.

We here investigate the irregularity of g-mode period spacings in a group of red giant branch (RGB) stars, which remains so far unexplained. High-radial order g modes are expected to be approximately equally spaced in period by $\Delta\Pi_l$, where $l$ is the mode degree. In red giants, dipole mixed modes can be used to measure $\dpun$ using asymptotic expressions of the mode frequencies (\citealt{mosser15}). While $\dpun$ is nearly constant over the frequency range of observed modes for the vast majority of RGB stars, some red giants show significant variations of $\dpun$ (\citealt{mosser18}, \citealt{deheuvels22}). We here show that this feature is the signature of strong magnetic fields in the cores of these stars. This constitutes a new way of detecting and 
characterizing magnetic field in the cores of red giants. 

In Sect. \ref{sect_dpun}, we present red giants that exhibit deviations from the regular period spacing pattern of g modes, and we find additional such targets in \kepler\ data. We then show in Sect. \ref{sect_mag_perturbations} that strong core magnetic fields can account for this phenomenon. In Sect. \ref{sect_mag_strength}, we determine the field strengths that are required to match the seismic observations. We discuss these measurements in Sect. \ref{sect_discussion}, before concluding in Sect. \ref{sect_conclusion}.

\section{Red giants with non-constant $\dpun$ \label{sect_dpun}}

\subsection{Previous detections of non-constant $\dpun$ in RGB stars \label{sect_previous}}

To first order, high-radial-order gravity modes are expected to be equally spaced in period. 
Among the 160 RGB stars studied by \cite{mosser18}, only one shows clear deviations from a regular period spacing of g modes (KIC3216736). The authors attributed this irregularity to a buoyancy glitch (that is, a sharp variation in the \vaisala\ frequency $N$), which induces periodic variations in the asymptotic period spacing $\dpun$.

More recently, \cite{deheuvels22} identified additional RGB stars with non-constant $\dpun$, in a different context. These stars appeared among a peculiar class of RGB stars that are located below the so-called ``degeneracy sequence'' in the $(\dn,\dpun)$ plane, where RGB stars regroup when electron degeneracy becomes strong in their core.
Most of the stars in this class are intermediate-mass stars and 
are thought to result from mass transfer (\citealt{deheuvels22}). The only four lower-mass stars with too-low $\dpun$ must have a different origin. Contrary to intermediate-mass stars, they all
show clear departures from a constant period spacing of g modes. Interestingly, the star identified by \cite{mosser18} (KIC3216736) is among these targets. This suggests that there might be a link between the non-constancy of $\dpun$ and the fact that its measured value is abnormally low. These four stars show only one detected mode per rotational multiplet.

\subsection{Additional targets \label{sect_add}}

We searched for other targets showing non-constant $\dpun$ among RGB stars with detected oscillations using the catalog of \cite{yu18}. To estimate the period spacings of g modes using dipole mixed modes, we computed the so-called ``stretched'' periods $\tau$, defined by the differential equation $\hbox{d}\tau = \hbox{d}P/\zeta$ (\citealt{mosser15}), where $\zeta$ corresponds to the fraction of the mode kinetic energy that is enclosed in the g-mode cavity ($\zeta$ tends to 1 for pure g modes, and 0 for pure p modes). When building \'echelle diagrams of these stretched periods, mixed modes are expected to align in a vertical ridge if $\dpun$ is constant and deviations from a regular period spacing induce curvature in this ridge.

\begin{figure}
\begin{center}
\includegraphics[width=0.49\linewidth]{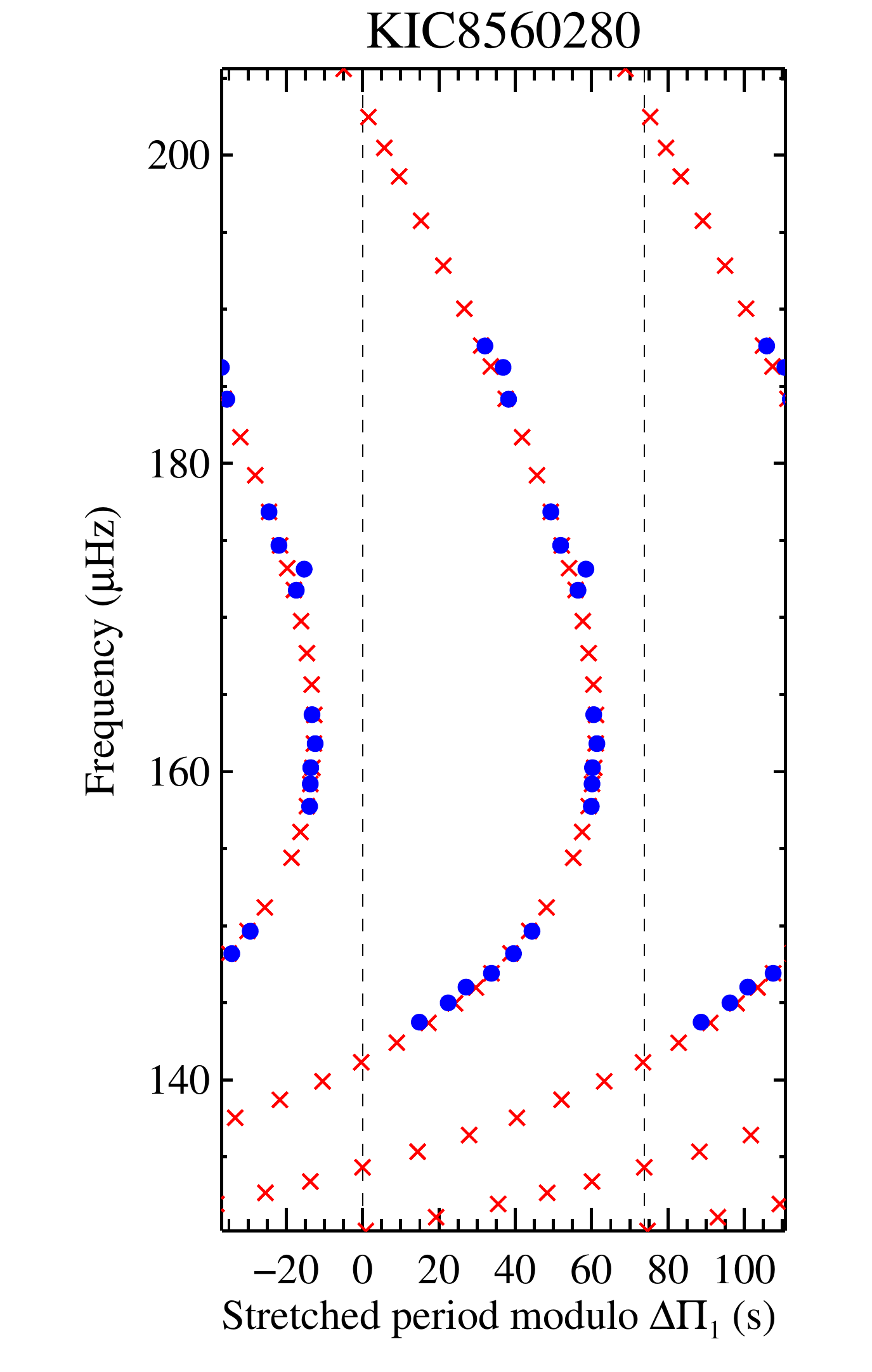}
\includegraphics[width=0.49\linewidth]{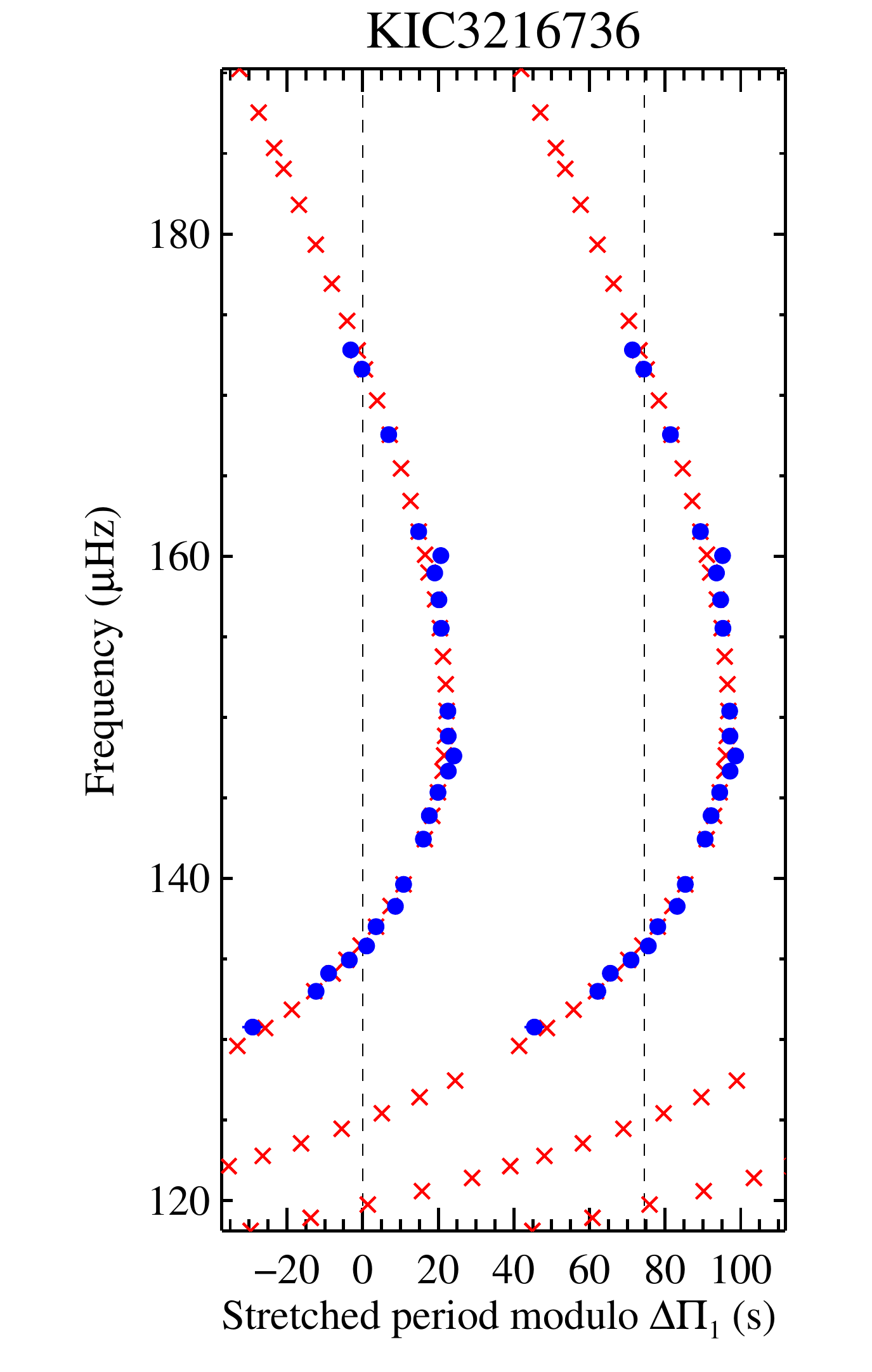}
\end{center}
\caption{Stretched period \'echelle diagrams of two red giants showing distortion from the regular g-mode pattern. Blue circles show detected dipole modes. Red crosses correspond to the best-fit asymptotic mixed mode frequencies obtained by including a magnetic perturbation.
\label{fig_stretch_vect}}
\end{figure}

We searched for stars with only one curved ridge detected in order to avoid the additional complication coming from rotational effects (these effects will be addressed in a subsequent work). This can mean that these stars are seen pole-on, so that only the $m=0$ modes can be detected. This could also arise if the core rotation is too weak to produce detectable rotational splitting in \kepler\ data.
We thus found seven additional targets, bringing the total of the sample to 11 stars (see Table \ref{tab_curved_RGB}). Their stretched period \'echelle diagrams are shown in Fig. \ref{fig_stretch_vect} and \ref{fig_stretch_vect_app}. They were folded using an average value of the asymptotic period spacing over the frequency range of the observations, which is further referred to as $\dpun^{\rm (meas)}$.

\begin{figure}
\begin{center}
\includegraphics[width=\linewidth]{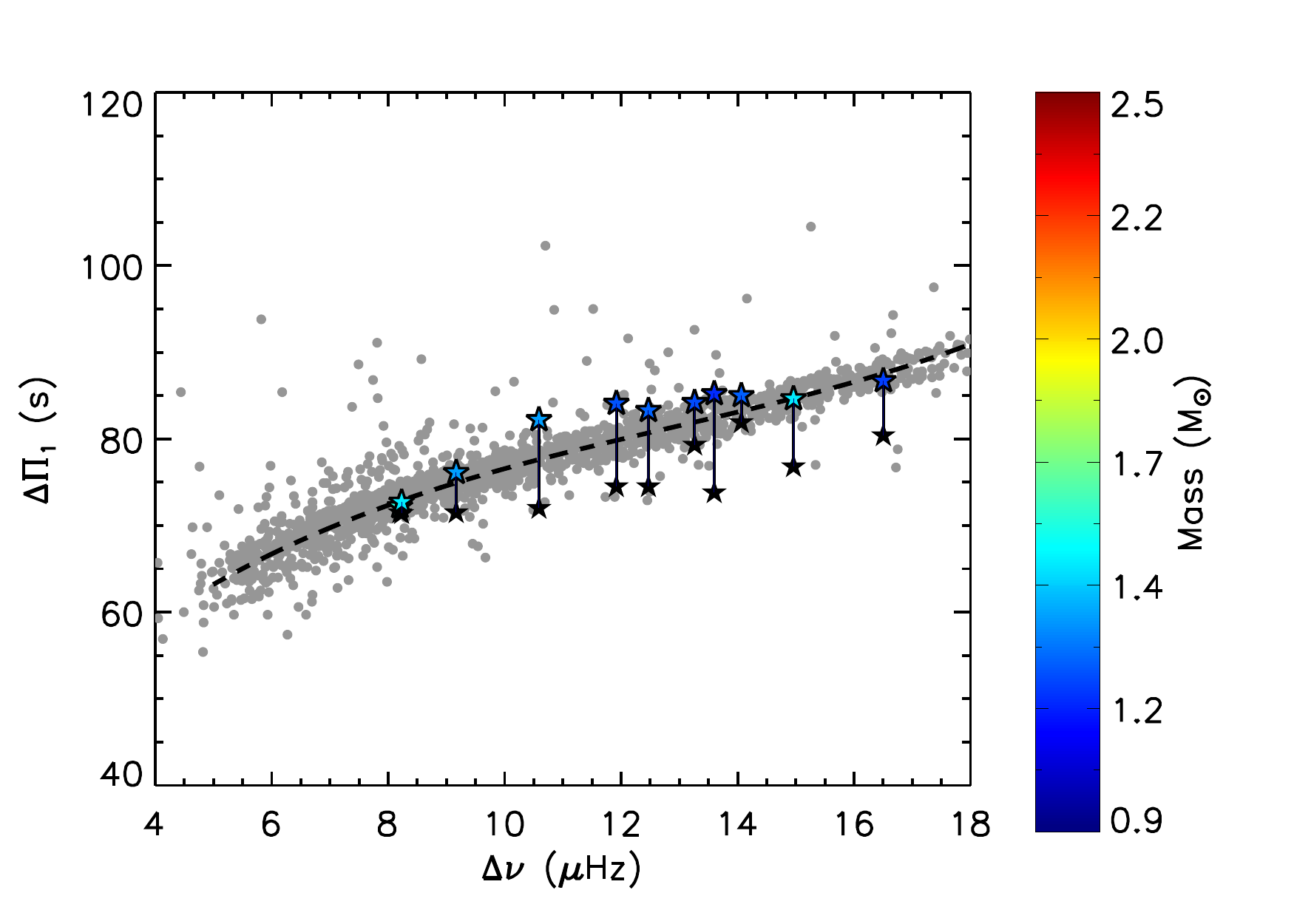}
\end{center}
\caption{Location of RGB stars with non-constant $\dpun$ in the $(\dn,\dpun)$ plane. Black star symbols correspond to the values of $\dpun$ that produce the best vertical alignment of modes in the stretched period \'echelle diagram (Sect. \ref{sect_dpun}). Colored star symbols correspond to the corrected values of $\dpun$ obtained by taking into account the magnetic perturbation to the mode frequencies (Sect. \ref{sect_mag_strength}). Other RGB stars from \cite{vrard16} are show as grey circles (for clarity, stars flagged by the authors as potential aliases were omitted).
\label{fig_deg_mag}}
\end{figure}

The location of these 11 targets in the $(\dn,\dpun)$ plane is shown in Fig. \ref{fig_deg_mag} (black star symbols), where we have used the values $\dpun^{\rm (meas)}$ as the measured asymptotic period spacing. Three of the seven additional targets lie well below the degeneracy sequence of RGB stars, which confirms the link between non-constant $\dpun$ and low measured values for these quantities.

\subsection{Origin of distortions in g-mode pattern}

Deviations from a regular $\dpun$ are generally attributed to buoyancy glitches (e.g., \citealt{mosser15}). To produce the observed distortions in $\dpun$, we showed in Appendix \ref{app_glitch} that a buoyancy glitch needs to have a large amplitude (the local value of $N$ must be multiplied by a factor of at least six) and be located either deep inside the inert He core, or well above the H-burning shell. While this cannot be excluded, no known process is expected to produce such strong features in these regions of an RGB star. Secondly, the shape of the modulation in $\dpun$ that is produced by a large-amplitude glitch strongly differs from the observations (see Appendix \ref{app_glitch}). Finally, the hypothesis of a buoyancy glitch would not explain why the measured values of $\dpun$ are unexpectedly low for most of our stars. It thus seems unlikely that the observed deviations in $\dpun$ arise from buoyancy glitches. In the following sections, we explore the possibility that the irregularities in $\dpun$ are produced by internal magnetic fields.

\section{Effects of magnetic fields on g-mode period spacings \label{sect_mag_perturbations}}

The influence of magnetic fields over oscillation mode frequencies has been studied over the last decades using a perturbative approach (\citealt{unno89}, \citealt{gough90}). Recently, their effects on mixed modes in red giants have been addressed in the special case of dipolar fields with specific radial profiles, either aligned with the rotation axis (\citealt{hasan05}, \citealt{gomes20}, \citealt{mathis21}, \citealt{bugnet21}) or inclined (\citealt{loi21}). 

\cite{li22} extended these studies to an arbitrary  magnetic field and obtained a general expression for the magnetic frequency shift that is valid provided that the azimuthal component is not much larger than the radial one ($B_{\phi}/B_r \ll \omega_{\rm max}/N$, where $\omega_{\rm max}$ is the angular frequency at the maximum power of oscillations and $N$ is the \vaisala\ frequency\footnote{The ratio $\omega_{\rm max}/N$ typically exceeds $10^2$ for red giants.}). Accordingly,
the multiplets of $l=1$ pure g modes (that is, g modes that are not coupled to p modes) undergo an average shift $\omega_{\rm B}$, given by
\begin{equation}
\omega_{\rm B} = \frac{\mathcal{I}}{\mu_0\omega^3} \int_{\ri}^{\ro} K(r) \overline{B_r^2} \hbox{d}r,
\label{eq_omegaB}
\end{equation}
where $K(r)$ is a weight function that probes the g-mode cavity and sharply peaks in the vicinity of the H-burning shell (HBS), $\mathcal{I}$ is a factor that depends on the core structure (see Eq. 45 and 46 of \citealt{li22}), and $\overline{B_r^2} = (4\pi)^{-1}\iint B_r^2 \sin\theta\,\hbox{d}\theta\,\hbox{d}\phi$. The angular frequency shifts of the components of g-mode dipole multiplets are then given by
\begin{align}
\domg(m=0) & = \left( 1-a \right) \omega_{\rm B} \label{eq_shift_m0} \\
\domg(m=\pm1) & = \left(1+\displaystyle\frac{a}{2}\right) \omega_{\rm B},
\label{eq_shift_m1}
\end{align}
where $a$ is a dimensionless coefficient that depends on the horizontal geometry of $B_r^2$ ($a \;\propto
\iint B_r^2 P_2(\cos\theta) \sin\theta \,\hbox{d}\theta\hbox{d}\phi$, where $P_2(\cos\theta)$ is the second order Legendre polynomial).

The dependency of magnetic shifts with $\omega^{-3}$ shows that low-frequency (that is, high-radial-order) g modes are more affected by magnetic fields. For this reason, magnetic shifts create a deviation from the regular period spacing of pure g modes, as was already pointed out by \cite{loi20b}, \cite{bugnet21}, and \cite{li22}. Very recently \cite{bugnet22} proposed a method to detect the signature of magnetic fields exploiting this property. 

For illustration, Fig. \ref{fig_stretch_vect} shows stretched \'echelle diagrams of mixed modes with magnetic perturbations (see Sect. \ref{sect_mag_strength}).
The left panel corresponds to a case where the unperturbed g modes have an asymptotic period spacing of $\dpun = 85$~s, and we have added a magnetic perturbation corresponding to a frequency shift of $3.9\,\mu$Hz at $\numax$. The ridge appears strongly curved, similarly to the observations. To visualize the ridge properly, the stretched \'echelle diagram was folded using a period spacing of $\dpun^{\rm (meas)} = 73.8$~s, which is much lower than the unperturbed period spacing $\dpun$.

Magnetic perturbations thus account for both characteristics of the stars identified in Sect. \ref{sect_dpun}. First, they produce curved ridges in the period \'echelle diagram. Since magnetic shifts are always positive, the period spacings of g modes decrease with decreasing mode frequency. Thus, the curvature always has the same shape, the low-frequency part of the ridge being bent to the left direction of the period \'echelle diagram. Interestingly, all the targets identified in Sect. \ref{sect_dpun} show ridges curved in this direction (see Fig. \ref{fig_stretch_vect_app}). Secondly, magnetic perturbations yield a measured period spacing that is significantly lower than the asymptotic unperturbed period spacing $\dpun$. This can explain why most of the targets identified in Sect. \ref{sect_dpun} are located below the degenerate sequence in the $(\dn, \dpun)$ plane.

\section{Measurement of magnetic field strengths \label{sect_mag_strength}}

We then estimated the field strengths that are required to account for the observations. For this purpose, we computed asymptotic expressions of the mixed mode frequencies including magnetic perturbations. We followed the method that we proposed in \cite{li22}, which is briefly recalled here.
The effects magnetic fields are taken into account by adding a magnetic perturbation to the frequencies of pure p and g modes. These perturbed frequencies are then plugged into the asymptotic expression of mixed mode frequencies given by \cite{shibahashi79}.
While the frequencies of p modes are unaffected (\citealt{li22}), the periods of g modes are expressed as
\begin{equation}
\pg = P_{\rm g,0} \left(1 + \frac{\domg}{2\pi} P_{\rm g,0} \right)^{-1} 
\end{equation}
where $P_{\rm g,0} = (n_{\rm g}+1/2+\epsg)\dpun$ is the first-order asymptotic expression of $l=1$ g modes without perturbation, and $\domg$ is the magnetic perturbation to g-mode frequencies. Using Eq. \ref{eq_omegaB}-\ref{eq_shift_m1}, $\domg$ can be written as $\domg = \delta\omega_0\left(\omega_{\rm max}/\omega\right)^3,$
where $\delta\omega_0$ corresponds to the magnetic shift at $\omega_{\rm max}$.

For the 11 stars of our sample, we optimized the values of $\dpun$, $\delta\omega_0$, $\varepsilon_{\rm g}$, and $d_{01}$ (defined below) to match the observations at best using a Markov chain Monte Carlo approach.
Based on the measurements of $\varepsilon_{\rm g}$ for hundreds of \kepler\ red giants by \cite{mosser18}, we assumed a gaussian prior on $\varepsilon_{\rm g}$ with a mean of 0.28 and a standard deviation of 0.08, and we considered uniform priors for the other parameters. 
The characteristics of pure p modes were derived from the observed radial modes, with the exception of $d_{01}$, defined as the average small separation $\nu_{{\rm p},l=0}-\nu_{{\rm p},l=1}+\dn/2$, which was considered as a free parameter of the fit.

 The optimal parameters of the fit are given in Table \ref{tab_curved_RGB}. The corresponding asymptotic frequencies are shown as red crosses in Fig. \ref{fig_stretch_vect} and \ref{fig_stretch_vect_app}. The agreement with the observations is very good, the curvature of the ridge being well reproduced for all the stars. The fit also provides an estimate of the unperturbed asymptotic period spacing $\dpun$ for these stars, which is, as expected, larger than the apparent period spacing $\dpun^{\rm (meas)}$. We used the newly determined values of $\dpun$ to update the location of the 11 targets in the $(\dn,\dpun)$ plane in Fig. \ref{fig_deg_mag} (colored star symbols). It is striking to observe that they now lie on the degenerate sequence, as expected for stars in this mass range and evolutionary state. Thus, there is a body of evidence that the distortions to the g-mode pattern that are observed in the 11 targets of the sample are indeed produced by internal magnetic fields. This yields the opportunity to characterize these fields.

The measurement of $\delta\omega_0$ can be used to derive an estimate of $\langle B_r^2 \rangle = \int_{\ri}^{\ro} K(r) \overline{B_r^2} \hbox{d}r$ using Eq. \ref{eq_omegaB}-\ref{eq_shift_m1}. The obtained expression depends on the asymmetry parameter $a$, which can unfortunately not be measured with only one component detected per multiplet. However, we have shown that $-1/2\leqslant a \leqslant 1$ (\citealt{li22}), so that we can place a lower limit on the value of $\langle B_r^2 \rangle$. We obtain
\begin{equation}
\langle B_r^2 \rangle_{\rm min} =  \frac{2}{3} \frac{\delta\omega_0 \omega_{\rm max}^3 \mu_0 }{ \mathcal{I}},
\label{eq_br2min}
\end{equation}
where $\mu_0$ is the magnetic permeability. This expression is valid regardless of whether the observed modes have an azimuthal number of $m=0$ or $m=\pm1$ (indeed, the factors $1-a$ and $1+a/2$ appearing in Eq. \ref{eq_shift_m0} and \ref{eq_shift_m1}, respectively, are both always inferior to $3/2$). Only in the very specific case of a field that is entirely concentrated on the poles ($a\rightarrow 1$) would the measured field be much larger than $\langle B_r^2 \rangle_{\rm min}$. For instance, if $B_r^2$ has an axisymmetric dipolar configuration ($a = 2/5$), we have $\langle B_r^2 \rangle = 5/2 \langle B_r^2 \rangle_{\rm min}$. 

To calculate $\langle B_r^2 \rangle_{\rm min}$, the term $\mathcal{I}$ must be known, for which a model of the stellar internal structure is needed. For this purpose, we used a pre-computed grid of stellar models of red giants with various masses, metallicities, and evolutionary stages, built with the evolution code \mesa\ (\citealt{paxton11}). For each target, we selected models from the grid that simultaneously reproduce the asymptotic large separation of p modes $\dn$ and the asymptotic period spacing of dipole g modes $\dpun$. The models that satisfy this condition all give similar estimates of $\mathcal{I}$. We thus obtained measurements of $\langle B_r^2 \rangle_{\rm min}^{0.5}$ ranging from about 40~kG to about 610~kG (see Table \ref{tab_curved_RGB}). 

\begin{figure}
\begin{center}
\includegraphics[width=\linewidth]{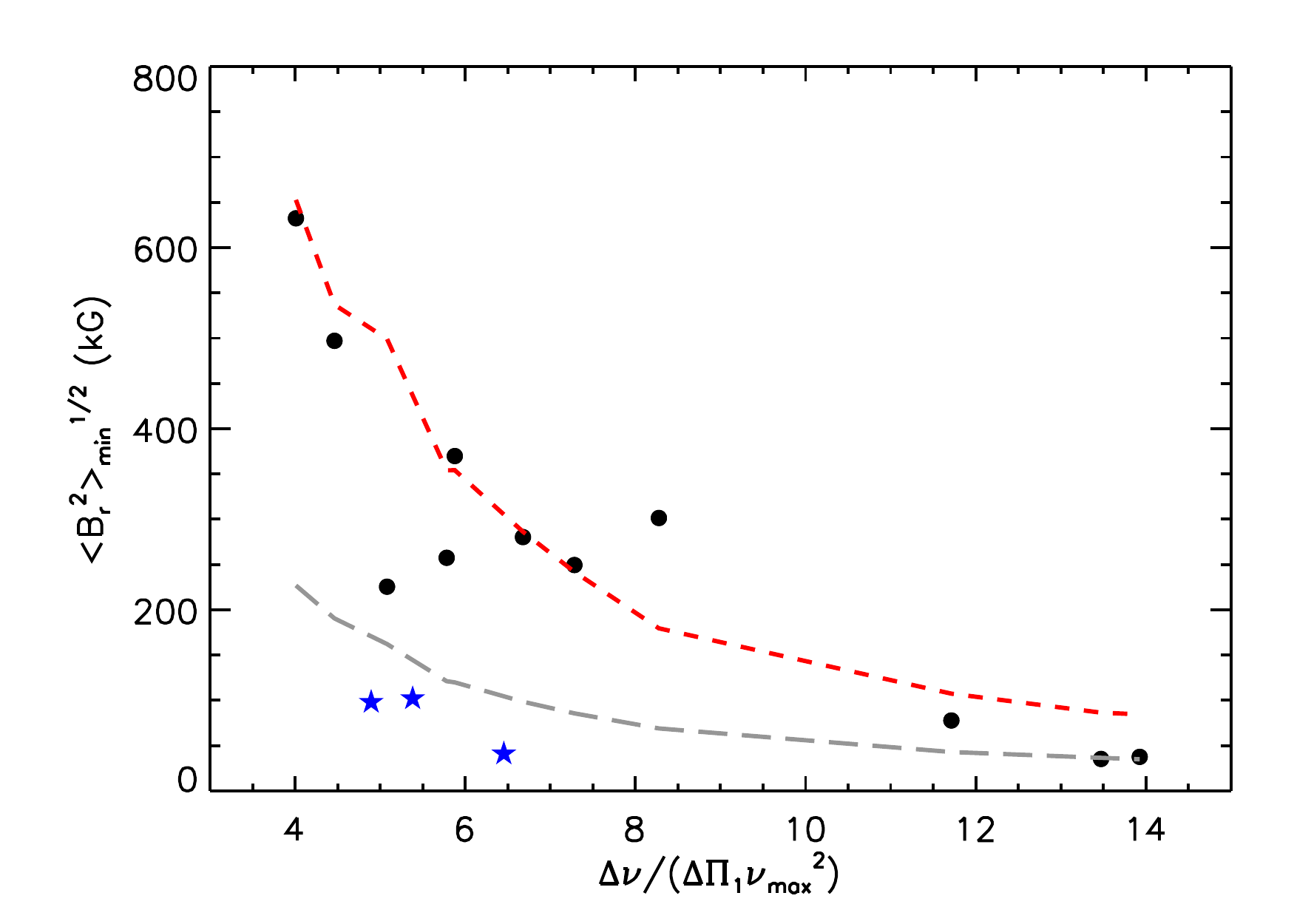}
\end{center}
\caption{Minimal field strength $\langle B_r^2 \rangle_{\rm min}^{0.5}$ required to account for the observed distortions in the g-mode period spacing for the 11 stars of our sample (black circles). They are plotted as function of the mixed mode density $\mathcal{N} = \dn/(\dpun\numax^2)$, which is a proxy for evolution along the RGB (\citealt{gehan18}).
The red dashed line indicates the critical field $B_{\rm c}$ and the grey long-dashed line shows the minimal field strength $B_{\rm th}$ required to detect the distortions in the g-mode pattern. The blue star symbols show the stars from \cite{li22}. 
\label{fig_calc_B}}
\end{figure}

\section{Discussion \label{sect_discussion}}

\subsection{Magnetic field strength vs evolution}

In Fig. \ref{fig_calc_B}, we plot the measured field strengths as a function of the density of mixed modes $\mathcal{N} = \dn/(\dpun\numax^2)$, which is a good proxy for the evolution along the red giant branch (\citealt{gehan18}). We observe a clear decrease of the measured field intensities along the evolution. At first sight, this trend is surprising. Indeed, assuming conservation of the magnetic flux, the contraction of the core as red giants evolve should increase the field intensity, so that one would have expected the opposite trend. Before interpreting this trend, we addressed the question of potential observational biases. In Appendix \ref{app_Bseuil}, we calculated the threshold field strength $B_{\rm th}$ that is required to produce detectable variations in the g-mode period spacing over the observed frequency range. As shown in Fig. \ref{fig_calc_B}, $B_{\rm th}$ decreases along the evolution on the RGB. This explains why we do not detect lower-intensity fields in unevolved red giants. However, the lack of higher-intensity fields in more evolved stars cannot be explained by this observational bias, and thus the decrease in the field strength with evolution seems real. 

\subsection{Comparison with the critical field}

We compared the measured minimal field intensities with the critical field $B_{\rm c}$. We stress that for fields over $B_{\rm c}$, a local analysis shows that gravity waves can no longer propagate (\citealt{fuller15}). While the details of how global modes are affected remain uncertain, it is clear that they will be impacted.
We used the stellar models selected from our grid in Sect. \ref{sect_mag_strength} to estimate $B_{\rm c}$ for each star of the sample. We evaluated $B_{\rm c}$ in the HBS, where it reaches a sharp minimum (\citealt{fuller15}), and where our field measurements have the highest sensitivity. Fig. \ref{fig_calc_B} shows that the value of $B_{\rm c}$ in the HBS decreases with evolution, as was already pointed out by \cite{fuller15}. We observe that our minimal field strength measurements closely follow the trend of $B_{\rm c}$ with evolution. One possible explanation for the trend observed in Fig. \ref{fig_calc_B} is that the core field increases with evolution, owing to magnetic flux conservation, and eventually reaches the critical field $B_{\rm c}$. Above this field, mixed modes would no longer form, making the seismic detection of core magnetic fields impossible.

\begin{figure}
\begin{center}
\includegraphics[width=\linewidth]{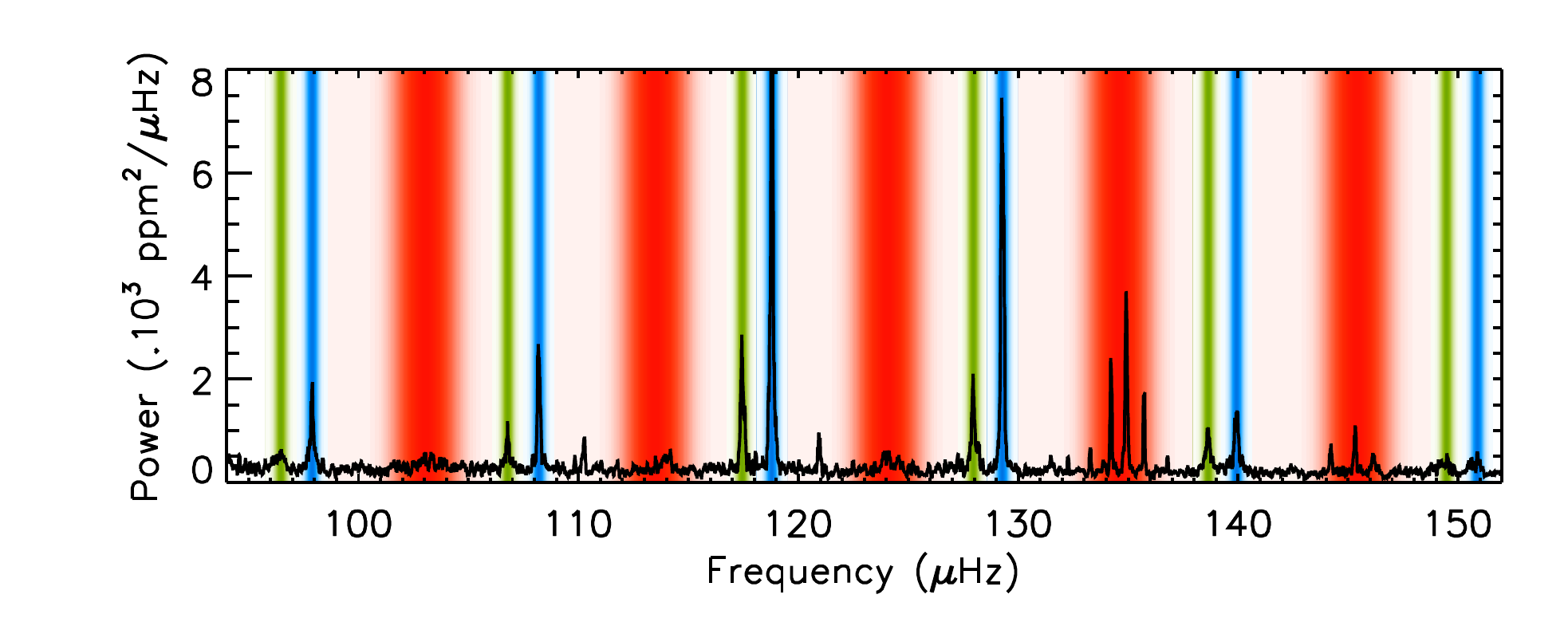}
\end{center}
\caption{Power spectrum of KIC\,6975038 obtained from \kepler\ data. Color-shaded areas indicate the location of $l=0$ (blue), $l=1$ (red) and $l=2$ (green) modes.
\label{fig_l1dep}}
\end{figure}

\subsection{Link with stars with suppressed dipole mixed modes \label{sect_l1dep}}

The ratio between the minimal measured field strength and the critical field $B_{\rm c}$ is maximal for KIC\,6975038, where it reaches a factor of about 1.7. Interestingly, this star shows clear signs of dipole mixed mode suppression. Fig. \ref{fig_l1dep} shows the power spectrum of KIC\,6975038 built with \kepler\ data. The regions of the spectrum where dipole mixed modes are expected are highlighted in red. While the dipole mixed mode pattern clearly appears at high frequency, it is nearly absent for frequencies around $\numax$ and below.

This type of behavior is expected, assuming that field intensities above the critical field $B_{\rm c}$ can suppress mixed modes. Indeed, $B_{\rm c}$ varies as $\omega^2$, so that for a given field strength, there exists a transition frequency $\omega_{\rm c}$ below which mixed modes should be strongly suppressed and above which they should be unaffected (\citealt{fuller15}, \citealt{loi20a}). This can be used to estimate the field strength for stars where the transition between suppressed and normal modes can be detected. For KIC6975038, the observed transition frequency $\omega_{\rm c}$ 
yields a radial field intensity of about 180 kG in the HBS. This estimate has the same order of magnitude as the minimal field strength $\langle B_r^2 \rangle_{\rm min}^{0.5} = 301$~kG that was inferred in an independent way using the perturbations to the g-mode period spacing (Sect. \ref{sect_mag_strength}). 
This star thus combines two different features that have been interpreted as potential indications of the presence of core magnetic fields and they both lead to comparable estimates of magnetic field strength. While more stars of this type would be required to draw conclusions, this is further indication that there might be a link between mixed mode suppression and strong core fields.

\subsection{Origin of the detected fields}

One possibility is that the detected fields were produced by a dynamo 
in the convective core during the main sequence. The stars of our sample have masses ranging from 1.11 to 1.56~$M_\odot$ (see Table \ref{tab_curved_RGB}). Contrary to the three stars studied in \cite{li22}, the lowest-mass stars likely had a radiative core during most the main sequence. However, even these stars possessed a small initial convective core at the beginning of the main sequence, owing to the burning of $^3$He and $^{12}$C outside of equilibrium (\citealt{deheuvels10}). The ohmic diffusion timescale being longer than the evolution timescale (\citealt{cantiello16}), these fields can have survived until the red giant phase and relaxed into stable configurations (\citealt{braithwaite04}). By using the stellar models introduced in Sect. \ref{sect_mag_strength} and assuming a conservation of the magnetic flux, we estimated the main-sequence field strengths that would be required to produce the detected fields (Appendix \ref{app_B_MS}). We found minimal field intensities ranging from 1 to 26~kG inside the main-sequence convective cores. This is in general lower than the radial magnetic field strengths found by the numerical simulation of a convective core (\citealt{brun05}) or order-of-magnitude estimates assuming equipartition with the convective motion kinetic energy (\citealt{cantiello16}). A dedicated study will be necessary to determine whether the measured core fields can be accounted for by this possible origin of the fields, taking into account the diversity of the dynamo-generated fields and the dissipation provoked by their relaxation (\citealt{becerra22}) and potential instabilities (\citealt{gouhier22}) in the post-main-sequence phase. 

\section{Conclusion \label{sect_conclusion}}

We here revisited the puzzling case of H-shell burning red giants that exhibit strong deviations from the regular period spacing that gravity modes should reach in the high-radial order limit (\citealt{deheuvels22}). We showed that this peculiarity is unlikely to be produced by buoyancy glitches, and on the contrary very well accounted for by strong magnetic fields in the core of these stars. We thus placed lower limits on the strength of the radial field in the vicinity of the H-burning shell, ranging from 40 to 610~kG for the 11 stars of our sample. We also showed that for one star, the measured field exceeds the critical field $B_{\rm c}$ above which gravity waves can no longer propagate in the core (\citealt{fuller15}). Interestingly, this star shows mixed mode suppression at low frequency, which further suggests that this phenomenon might be related to strong core magnetic fields, although it should be noted that the mechanisms leading to mode suppression remain uncertain. This study focused on red giants with one single component detected per multiplet, to avoid the additional complication arising from rotational effects. We plan to search more generally for similar behavior in \kepler\ data in the near future.

\begin{acknowledgements}
S.D., J.B. and F.L. acknowledge support from from the project BEAMING ANR-18-CE31-0001 of the French National Research Agency (ANR) and from the Centre National d'Etudes Spatiales (CNES).
\end{acknowledgements}

\bibliographystyle{aa.bst} % style aa.bst
\bibliography{biblio} % your references Yourfile.bib

\clearpage

\begin{appendix}

\section{Can the deviations in g-mode period spacings be related to buoyancy glitches? \label{app_glitch}}

It is well known that buoyancy glitches induce deviations in the pattern of high-radial-order g modes (e.g., \citealt{miglio08}). Such deviations were already found by exploiting the mixed modes of core-helium burning giants (\citealt{mosser15}). \cite{cunha15} and \cite{cunha19} provided the appropriate formalism to determine the properties of buoyancy glitches (location and amplitude) from their seismic signature. We here investigate what types of glitches could produce the strong deviations in the period spacings of g modes that we observed in our sample of \kepler\ red giants. 

A buoyancy glitch produces a periodic modulation in the period spacings of g modes. The period of this modulation is directly related to the position of the glitch. This position is generally expressed in terms of its buoyancy radius or depth. Following the notations of \cite{cunha19}, the buoyancy radius $\omega_{\rm g}^r$ and depth $\tilde{\omega}_{\rm g}^r$ at a radius $r$ are defined as
\begin{equation}
    \omega_{\rm g}^r = \int_{r_1}^{r} \frac{LN}{r}\,\hbox{d}r \;\;\; \hbox{;} \;\;\; \tilde{\omega}_{\rm g}^r = \int_{r}^{r_2} \frac{LN}{r}\,\hbox{d}r,
\end{equation}
respectively, where $L = [l(l+1)]^{1/2}$, and $r_1$ and $r_2$ are the inner and outer turning points of the g-mode cavity. We also introduce the total buoyancy radius of the g-mode cavity $\omega_{\rm g} \equiv \omega_{\rm g}^{r_2}$. For a glitch located at a radius $r^\star$, one period of the modulation covers $\Delta n$ radial orders, where
\begin{equation}
    \Delta n = \omega_{\rm g}/\omega_{\rm g}^{r\star},
    \label{eq_period_glitch}
\end{equation}
if the glitch is located in the inner half of the cavity ($\omega_{\rm g}^r\star < 0.5$). If it is located in the outer half, then $\omega_{\rm g}^{r\star}$ needs to be replaced by $\tilde{\omega}_{\rm g}^{r\star}$ in Eq. \ref{eq_period_glitch}. The amplitude of the modulation depends on the sharpness of the variations in $N$.

\subsection{Glitch location}

The g-mode period spacings can be obtained from the periods of mixed modes by applying a stretching (see Sect. \ref{sect_dpun}). The difference $\Delta\tau$ between the stretched periods of consecutive mixed modes (shown as an illustration for KIC5180345 in Fig. \ref{fig_glitch_mag}) provides an estimate of the g-mode period spacing. For all the stars in our sample, the observed deviations do not show the periodic behavior that is expected for buoyancy glitches. If the observed deviations arise from glitches, the period of the modulation needs to be larger than the range defined by the observed modes. For example, for KIC5180345 the glitch period would have to cover at least 40 radial orders. Using Eq. \ref{eq_period_glitch} and the stellar model of KIC5180345 obtained in Sect. \ref{sect_mag_strength}, this means that the buoyancy glitch would need to be located either very deep within the g-mode cavity (below a fractional radius of $10^{-4}$, that is, deep within the inert He core) or nearly at the outer edge of the g-mode cavity (that is, well above the H-burning shell). 

\subsection{Glitch amplitude}

\begin{figure}
\begin{center}
\includegraphics[width=\linewidth]{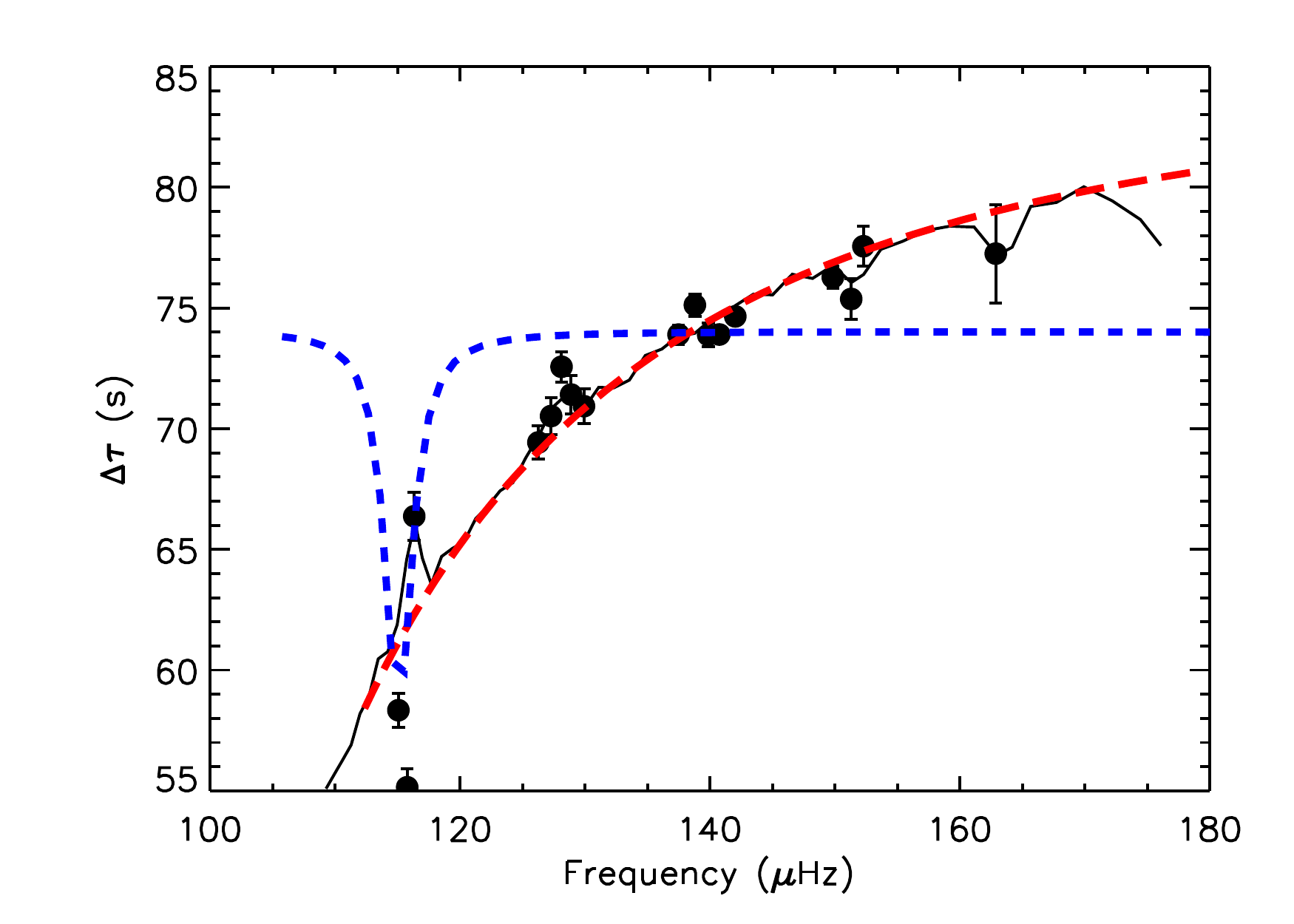}
\end{center}
\caption{Variations in the g-mode period spacings as a function of mode frequency for KIC5180345. The observed period spacings (filled circles) were computed as the difference $\Delta\tau$ between the stretched periods $\tau$ of consecutive dipolar mixed modes (see Sect. \ref{sect_add}). The blue dashed line indicates the g-mode period spacings for the best-fit buoyancy glitch perturbation. The red long-dashed line corresponds to the best-fit magnetic perturbation (see Sect. \ref{sect_mag_strength}). For the magnetic perturbation, we also show the differences $\Delta\tau$, which are directly comparable to the observations (black solid line).
\label{fig_glitch_mag}}
\end{figure}

We also addressed the question of the glitch amplitude that would be required to reproduce the observations. As shown in Fig. \ref{fig_glitch_mag}, the observed deviations have an amplitude that reaches about 30\% of the average period spacing. For comparison, \cite{cunha19} show the example of a Gaussian-shaped glitch in an RGB star, with an amplitude of about twice the local value of $N$ and a width of about $0.001\,R$ (see their Fig. 1). They find that it yields a modulation in the g-mode period spacing corresponding to only 1.5\% of the asymptotic period spacing (see their Fig. 5).

To roughly estimate the glitch amplitude that would be needed in our case, we used the formalism of \cite{cunha19}. We assumed a Gaussian-shaped buoyancy glitch and we used a Markov chain Monte Carlo (MCMC) to optimize the glitch properties (amplitude and width) in order to reproduce at best the observed g-mode period spacings. Fig. \ref{fig_glitch_mag} shows the best-fit solution (blue dashed line). This fitting problem appears to be highly degenerated: similar profiles may be generated by different sets of parameters. However, some properties of the glitch can be derived from the MCMC. In particular, we conclude that its amplitude must be greater than six times the local value of $N$. Any smaller value fails to reproduce the amplitude of the deviations observed in g-mode period spacings. 
However, even with the appropriate glitch amplitude, Fig. \ref{fig_glitch_mag} clearly shows that the best-fit solution cannot correctly reproduce the shape of the modulation. Indeed, it is well known \citep[e.g.,][]{miglio08,cunha19} that large-amplitude glitches yield modulations in the g-mode period spacings that involve sharp localized features (as opposed to small glitches, which produce sinusoidal modulations), which seem incompatible with the smoothly varying period spacings that are observed. On the contrary, a magnetic perturbation to the oscillation modes provides a very good agreement with the observations (red and black lines in Fig. \ref{fig_glitch_mag}). 

\section{Fit of asymptotic frequencies including magnetic perturbation}

\begin{figure*}
\begin{center}
\includegraphics[width=0.26\linewidth]{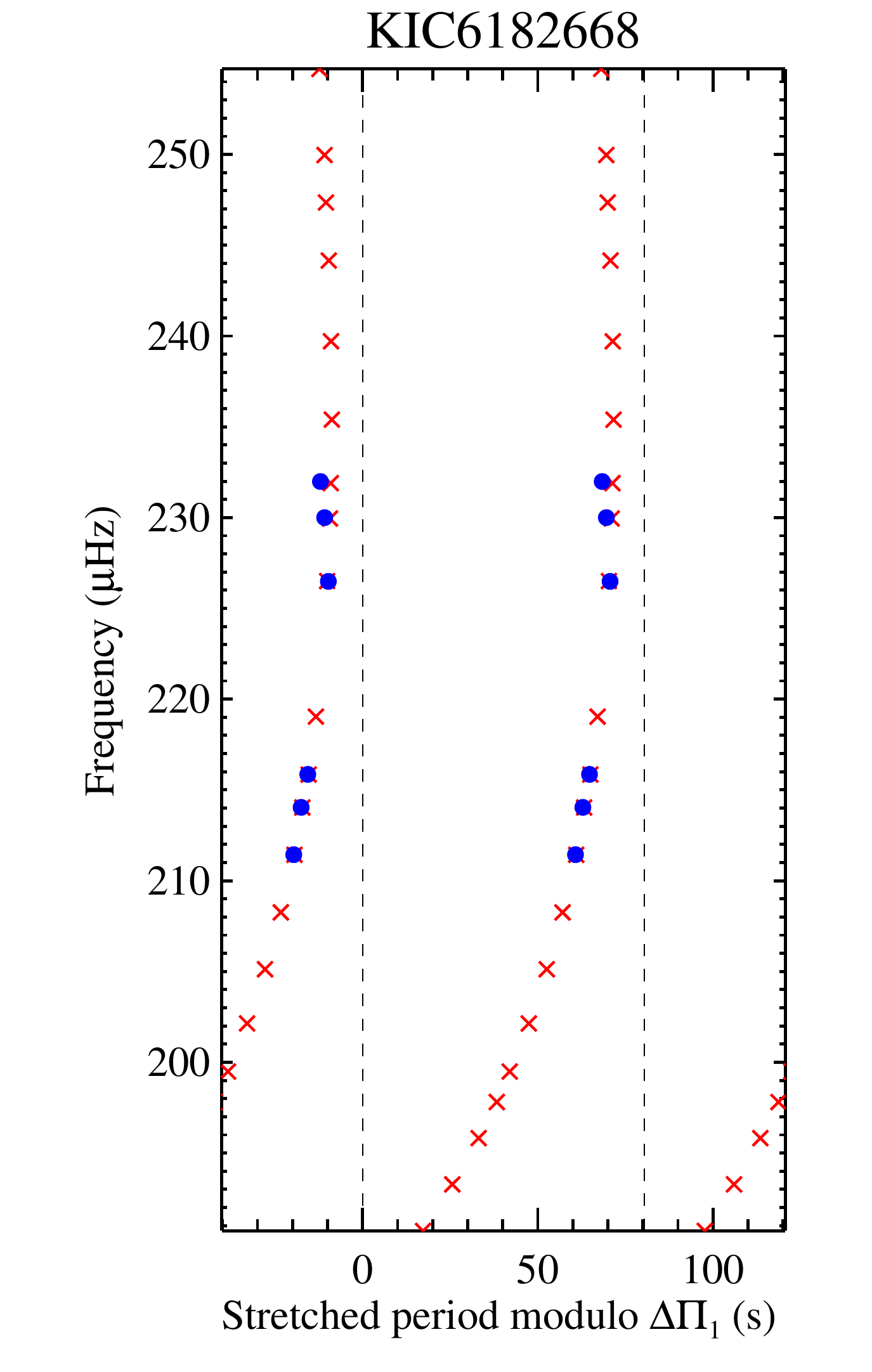}
\includegraphics[width=0.26\linewidth]{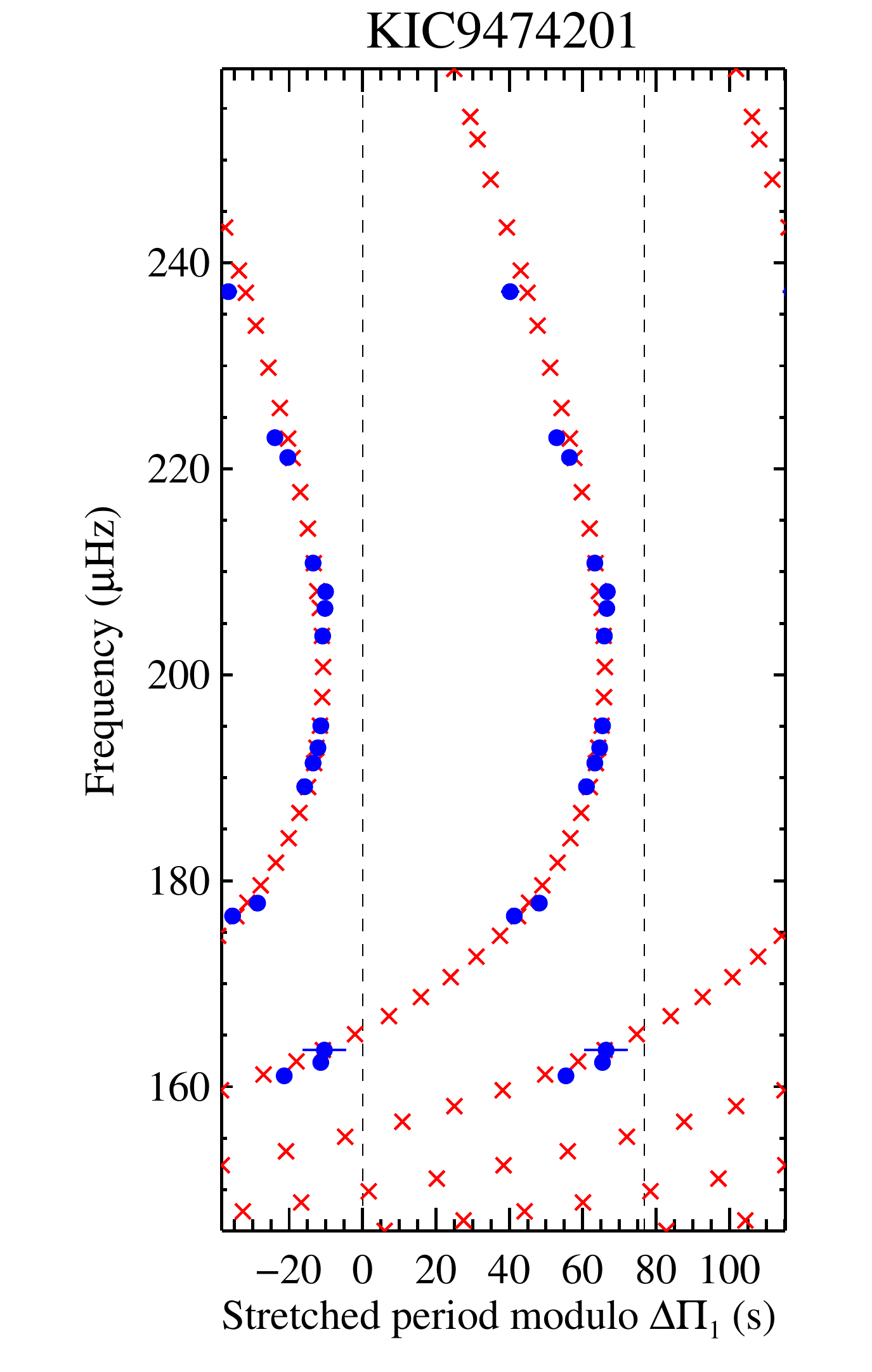}
\includegraphics[width=0.26\linewidth]{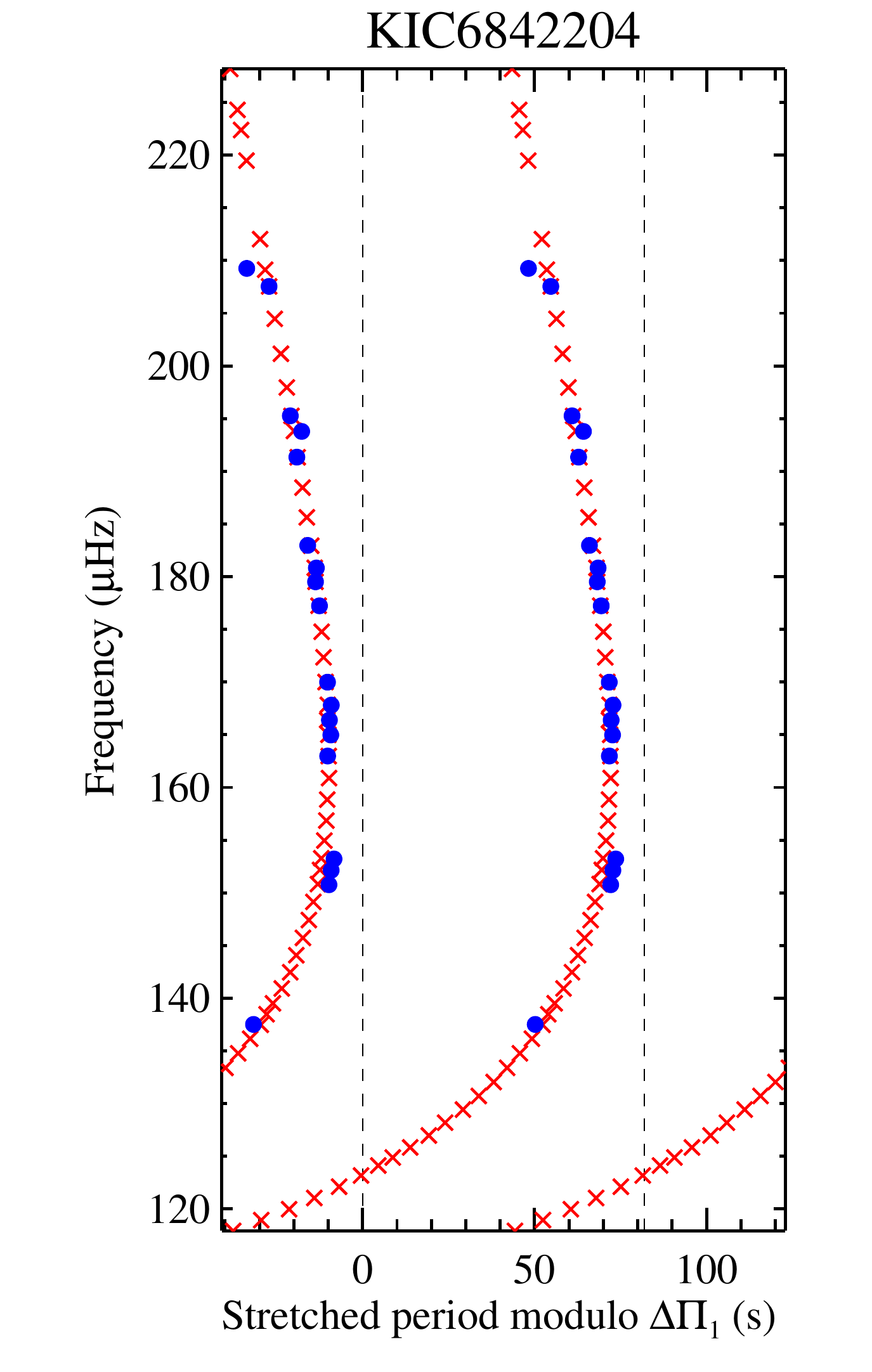}
\includegraphics[width=0.26\linewidth]{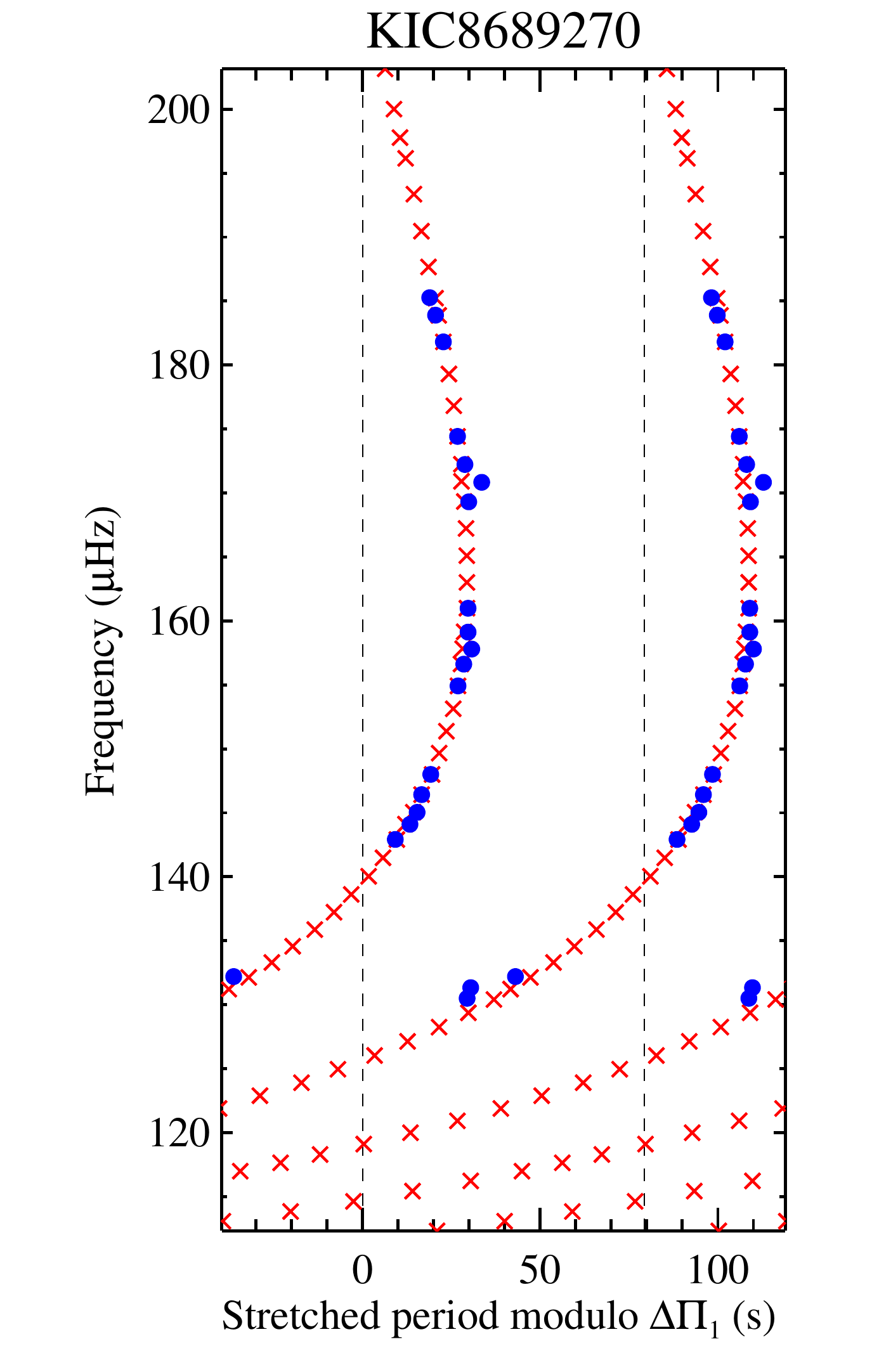}
\includegraphics[width=0.26\linewidth]{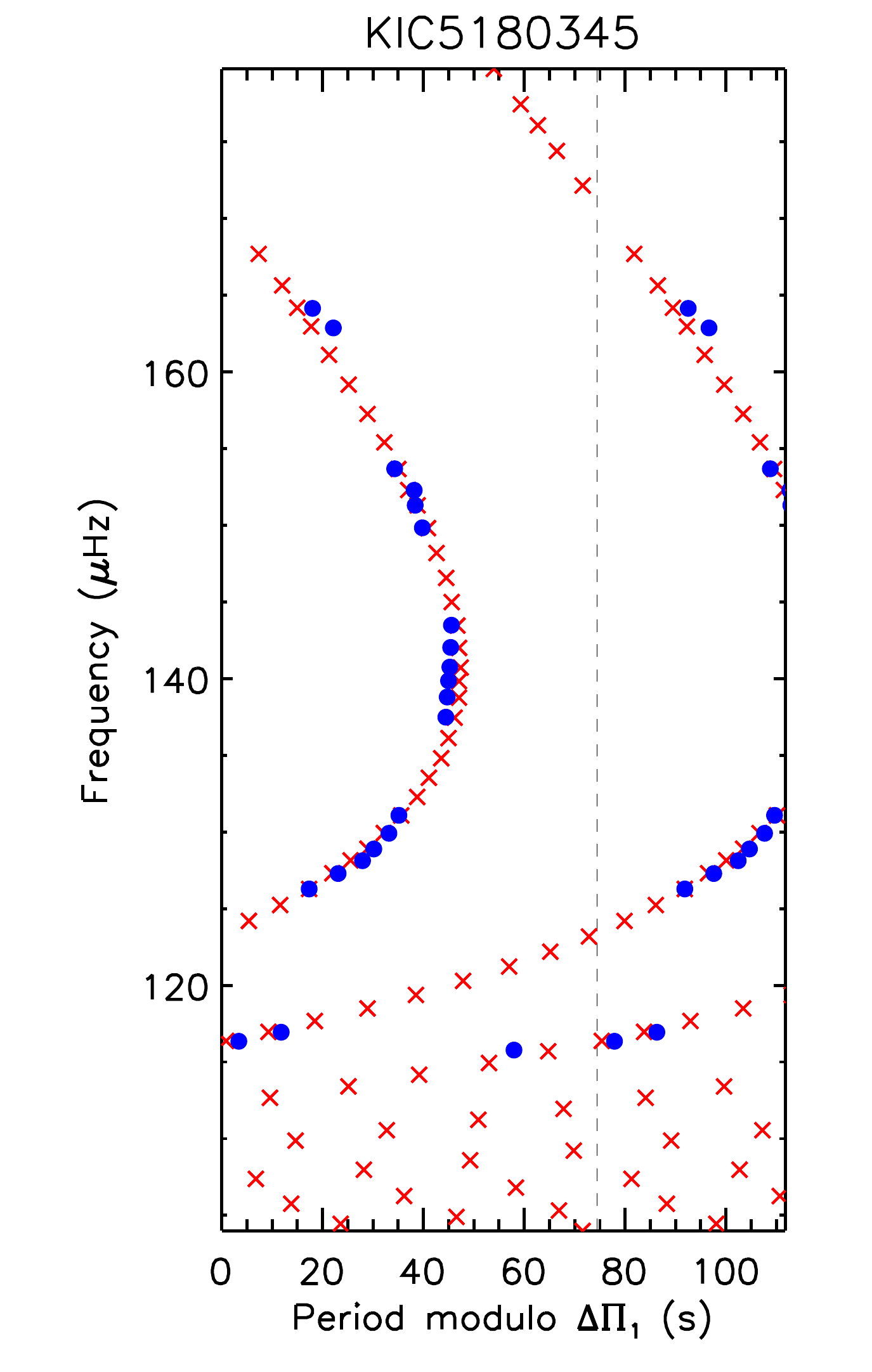}
\includegraphics[width=0.26\linewidth]{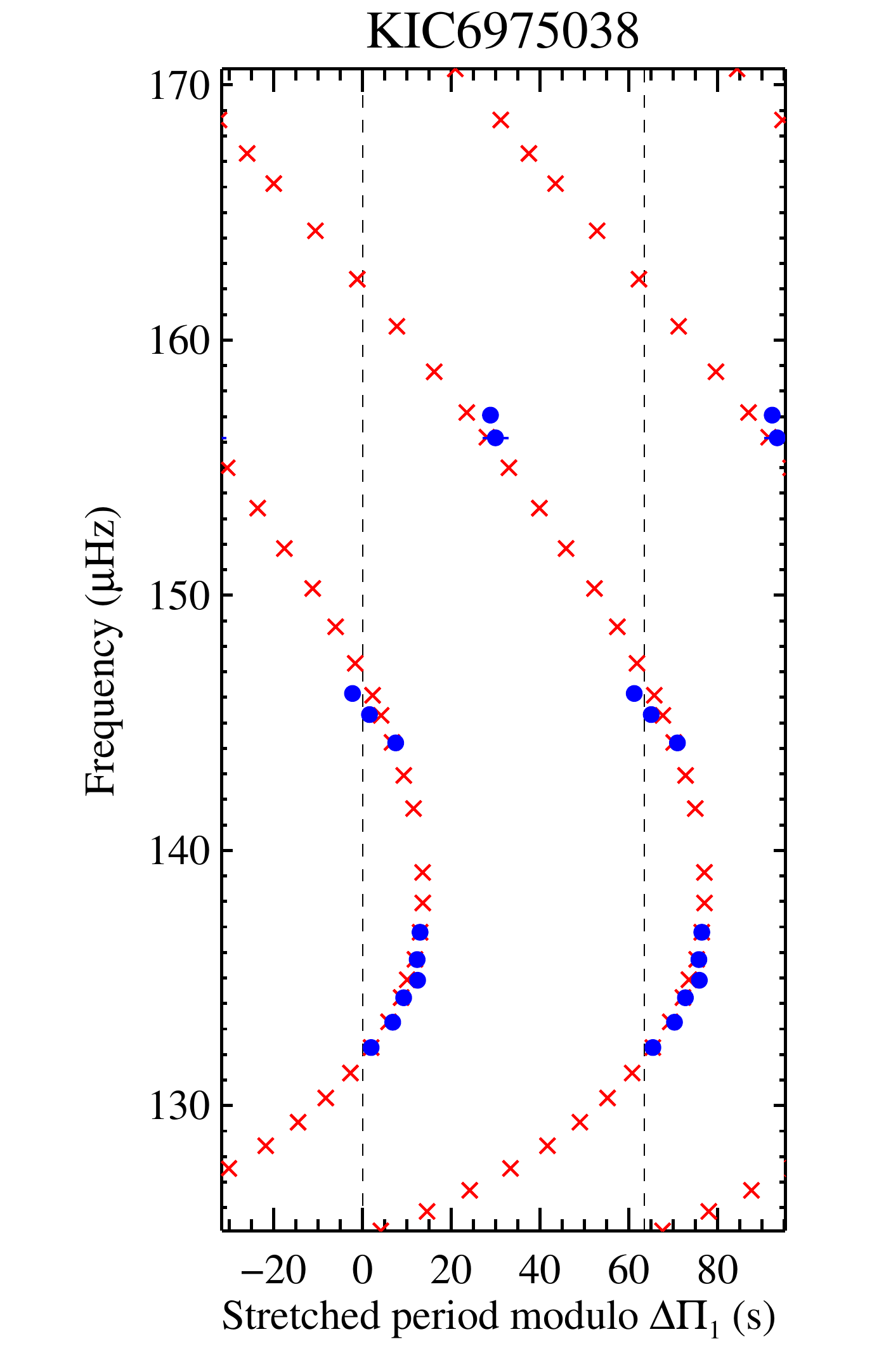}
\includegraphics[width=0.26\linewidth]{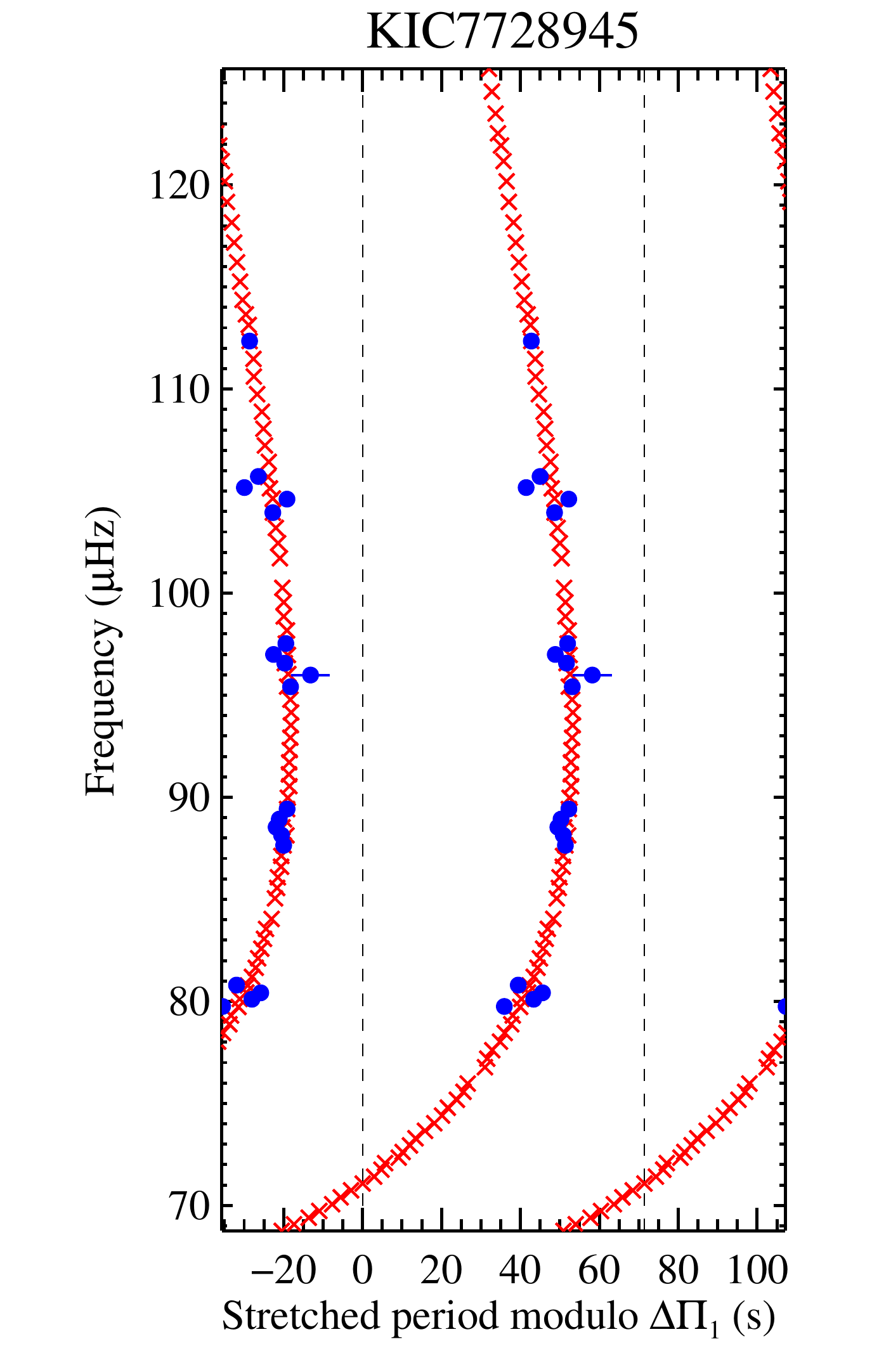}
\includegraphics[width=0.26\linewidth]{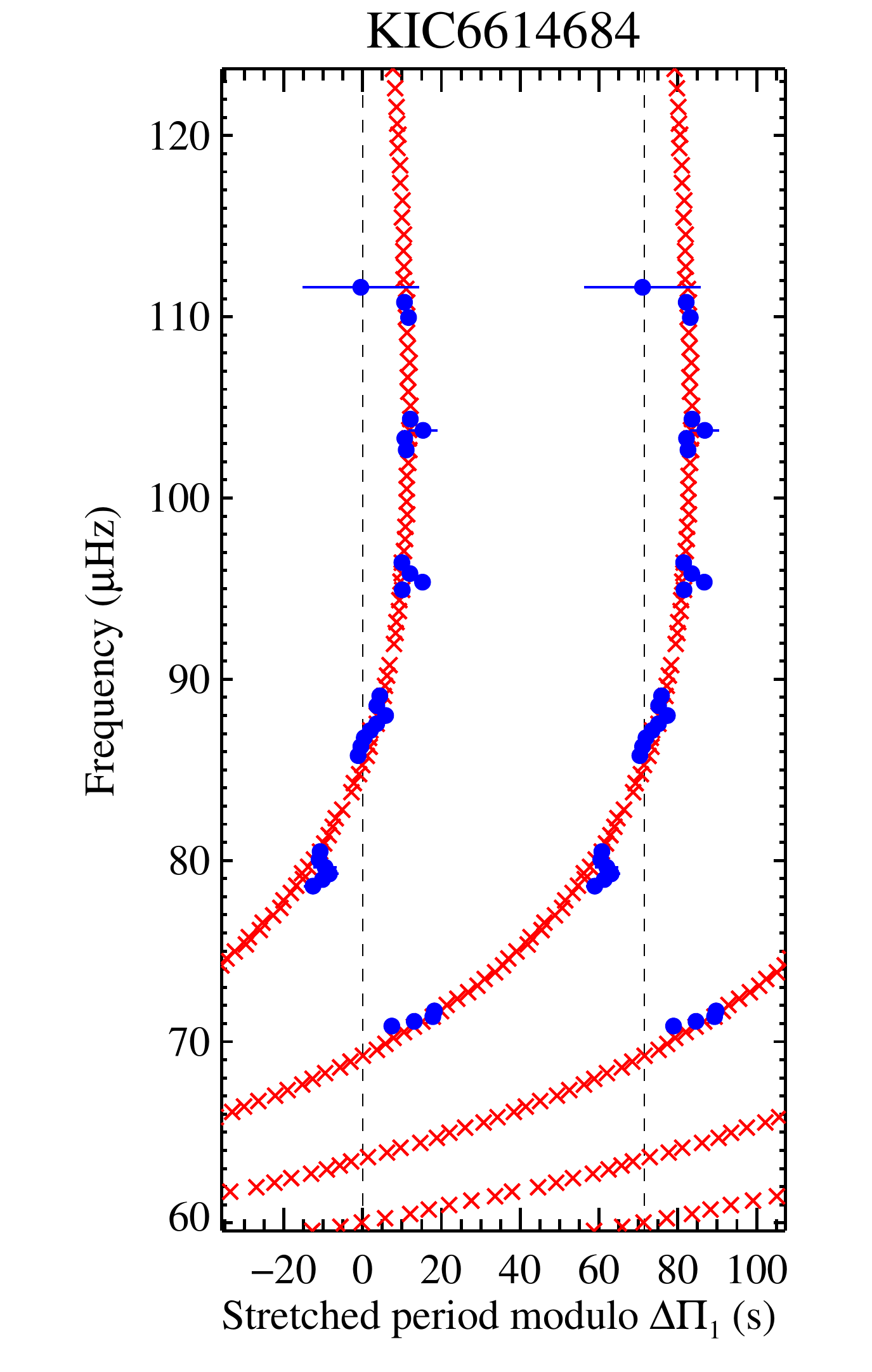}
\includegraphics[width=0.26\linewidth]{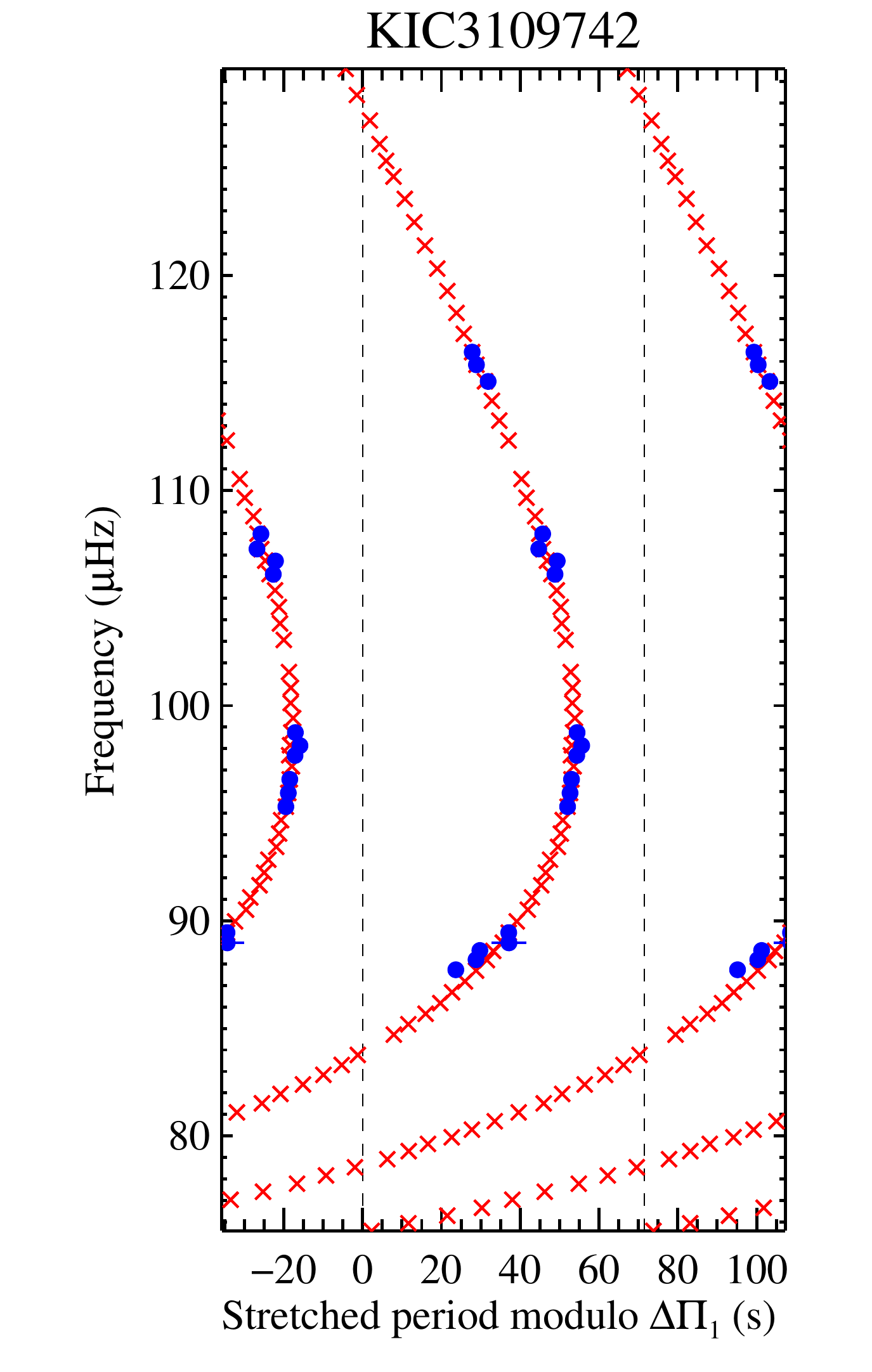}
\end{center}
\caption{Same as Fig. \ref{fig_stretch_vect} for the remaining stars of the sample.  
\label{fig_stretch_vect_app}}
\end{figure*}

Fig. \ref{fig_stretch_vect_app} shows the stretched \'echelle diagrams of the detected dipole mixed modes for 11 red giants in our sample (blue circles). They were folded using the apparent (perturbed) period spacing $\dpun^{\rm (meas)}$. In Sect. \ref{sect_mag_strength}, we fit an asymptotic expression of the mode frequencies including a magnetic perturbation to the observed modes. The optimal solutions are shown in Fig. \ref{fig_stretch_vect_app} as red crosses. The agreement is very good. We give in Table \ref{tab_curved_RGB} the parameters of the best-fit solutions, along with general stellar properties, for each star.

\begin{table*}
\begin{center}
\caption{Red giant branch stars showing strong variations in $\dpun$ among the catalog of \cite{yu18}. \label{tab_curved_RGB}}
\begin{tabular}{| l | c c c c c | c c | c |}
\hline
\T KIC Id & $\dn$ & $\numax$ & $M^{(a)}$ & $\mathcal{N}$ & $\dpun^{\rm (meas)}$ & $\dpun$ & $ \delta\omega_0/(2\pi)$ & $\langle B_r^2 \rangle_{\rm min}$ \\
\B & ($\mu$Hz) & ($\mu$Hz) & $(M_\odot)$ & & (s) & (s) & ($\mu$Hz) & kG \\
\hline
\T  6182668 & $16.48 \pm 0.12$ & $213.31 \pm  2.69$ & $1.24 \pm 0.09$ &  4.18 & 80.4 & $86.72 \pm  0.15$ & $ 4.49 \pm  0.09$ & 609 \\
     9474201 & $14.99 \pm 0.02$ & $197.58 \pm  1.41$ & $1.53 \pm 0.09$ &  4.54 & 76.8 & $84.65 \pm  0.07$ & $ 3.71 \pm  0.02$ & 488 \\
     6842204 & $14.14 \pm 0.04$ & $179.79 \pm 0.63$ & $1.26 \pm 0.06$ &  5.14 & 81.9 & $85.01 \pm  0.05$ & $ 0.87 \pm  0.02$ & 264 \\
    8560280 & $13.43 \pm 0.07$ & $165.82 \pm 0.75$ & $1.11 \pm 0.06$ &  5.73 & 73.8 & $85.23 \pm  0.05$ & $ 3.85 \pm  0.01$ & 369 \\
     8689270 & $13.21 \pm 0.03$ & $164.91 \pm 0.55$ & $1.14 \pm 0.05$ &  5.77 & 79.3 & $84.21 \pm  0.05$ & $ 1.83 \pm  0.01$ & 283 \\
    3216736 & $12.44 \pm 0.02$ & $150.38 \pm 0.71$ & $1.23 \pm 0.08$ &  6.61 & 74.5 & $83.29 \pm  0.03$ & $ 3.03 \pm  0.01$ & 278 \\
    5180345 & $11.81 \pm 0.03$ & $140.52 \pm 0.50$ & $1.17 \pm 0.07$ &  7.11 & 74.5 & $84.11 \pm  0.04$ & $ 3.07 \pm  0.01$ & 248 \\
    6975038 & $10.52 \pm 0.02$ & $127.98 \pm 0.80$ & $1.29 \pm 0.08$ &  7.81 & 63.5 & $82.22 \pm  0.08$ & $ 7.18 \pm  0.03$ & 286 \\
    3109742 & $ 9.09 \pm 0.01$ & $101.67 \pm 0.55$ & $1.32 \pm 0.09$ & 11.55 & 71.5 & $76.14 \pm  0.05$ & $ 1.07 \pm  0.01$ & 73 \\
    6614684 & $ 8.14 \pm 0.01$ & $92.04 \pm 0.45$ & $1.56 \pm 0.10$ & 13.31 & 71.5 & $72.19 \pm  0.01$ & $ 0.29 \pm  0.00$ & 38 \\
\B     7728945 & $ 8.27 \pm 0.02$ & $91.33 \pm 0.44$ & $1.51 \pm 0.09$ & 13.64 & 71.4 & $72.71 \pm  0.03$ & $ 0.35 \pm  0.01$ & 35 \\
\hline
\end{tabular}
\end{center}
\small{\textbf{Notes.} $^{(a)}$: Masses from \cite{yu18}.}
\end{table*}

\section{Minimal field strength required to detect magnetic distortion in g-mode pattern \label{app_Bseuil}}

We search for the minimal field strength that produces a detectable deviation in the regular period spacing of pure gravity modes. For this purpose, we consider typical oscillation properties for red giants. More refined estimates could be obtained on a star-to-star basis, but we are here interested in deriving broad estimates of magnetic intensity thresholds in order to investigate observational biases.

For a given red giant with a large separation $\dn$ and a frequency of maximum power of the oscillations $\numax$, we consider that the modes can be detected in a frequency interval ranging from $f_{\rm min} = \numax-2\dn$ and $f_{\rm max} = \numax+2\dn$. The asymptotic expression of unperturbed pure gravity modes is given by $P_n = \dpun(n+ 1/2 + \epsg)$, so that we expect to detect g modes with radial orders ranging from $n_{\rm min} = 1/(\dpun f_{\rm max})-\epsg-1/2$ and $n_{\rm max} = 1/(\dpun f_{\rm min})-\epsg-1/2$. We then consider the asymptotic periods $P'_n$ of perturbed g modes in the presence of a field that produces a frequency shift $\delta\nu_0$ at $\nu_{\rm max}$. Assuming that the perturbation remains small compared to the mode periods themselves (this is well verified at the detection limit for all stars of the sample), we have
\begin{equation}
    P'_n \approx P_n (1-\delta\nu_0 P_n^4 \numax^3).
\end{equation}

When analyzing the seismic data in red giants, high-radial order gravity modes are assumed to be regularly spaced in period and they are thus fit by an expression of the type $\overline{P'_n} = \Delta\Pi_1^{\rm (meas)}(n+ 1/2+ \varepsilon_{\rm g}^{\rm (meas)})$. Magnetic perturbations to the g-mode periods can be detected if the deviations compared to a regular spacing in period, expressed as $\delta P_n = \overline{P'_n} - P'_n$, are sufficiently large. Since the unperturbed periods $P_n$ vary linearly with $n$, the deviations $\delta P_n$ can be written as
\begin{equation}
    \delta P_n  = \left( \alpha n + \beta + P_n^5\numax^3 \right) \delta\nu_0 
    \label{eq_residuals}
\end{equation}
where $\alpha$ and $\beta$ are the parameters of a linear regression of the term $P_n^5\numax^3$ as a function of $n$. Eq. \ref{eq_residuals} shows that the intensity of the deviation from a regular period spacing is proportional to $\delta\nu_0$. This can also be seen in the measured period spacing $\Delta\Pi_1^{\rm (meas)}$, which here corresponds to $\Delta\Pi_1+\alpha\delta\nu_0$.

\begin{figure}[H]
\begin{center}
\includegraphics[width=\linewidth]{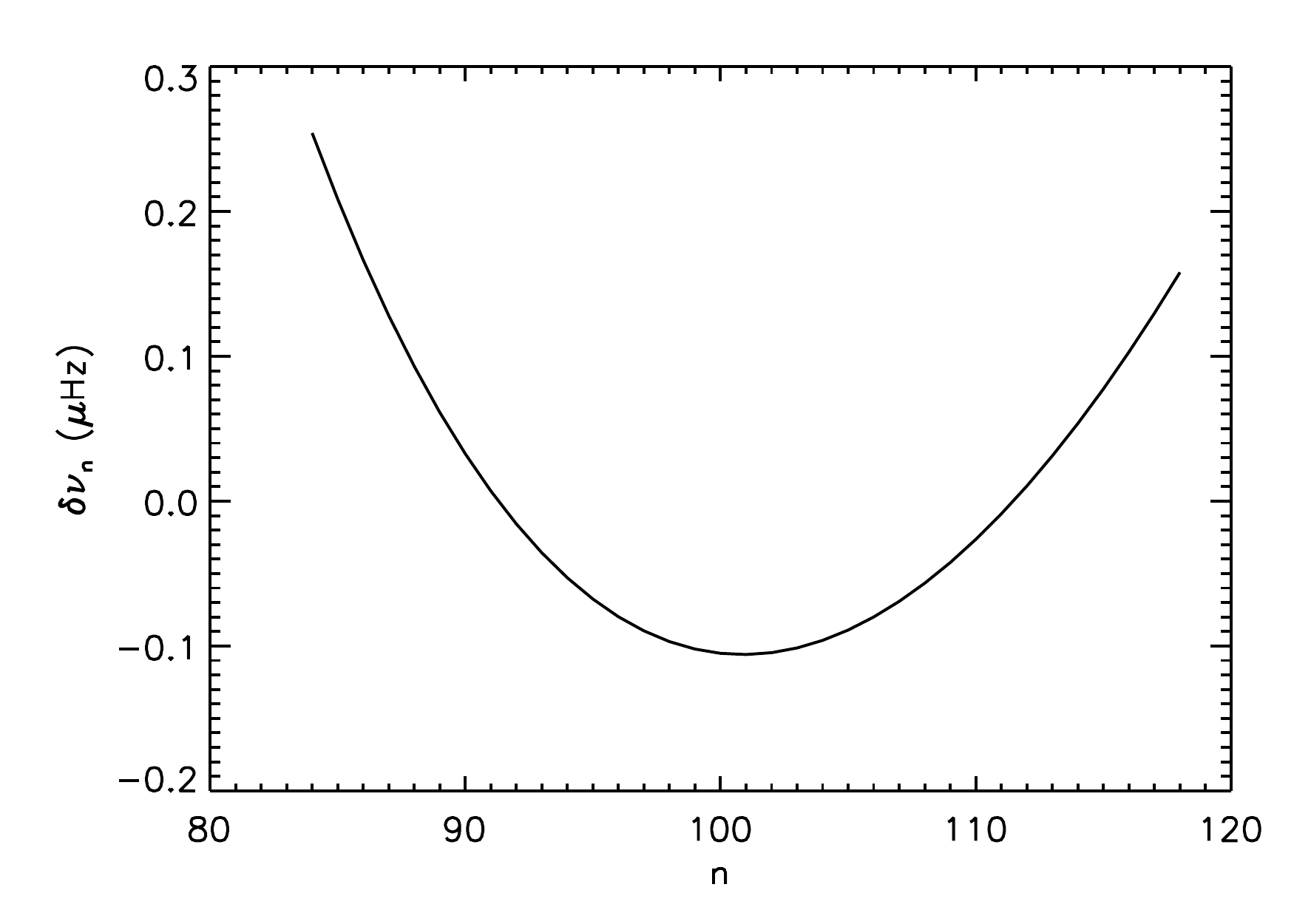}
\end{center}
\caption{Departures from a regular period spacing of gravity modes in the presence of a magnetic field, shown as frequency differences $\delta\nu_n$ as a function of the radial order $n$ of gravity modes.
\label{fig_deviations_mag}}
\end{figure}

The period differences can be translated into frequency differences as $\delta\nu = -\delta P_n/P_n^2$. Fig. \ref{fig_deviations_mag} shows the variations in $\delta\nu_n$ as a function of $n$ for an illustration case with $\dn = 10.6\,\mu$Hz, $\Delta\Pi_1 = 79.9$~s, $\varepsilon_{\rm g} = 0.3$, and $\delta\nu_0 = 0.4\,\mu$Hz. The maximal values of $|\delta\nu_n|$ are reached at the boundaries of the interval, more particularly for $n=\nmin$.

To determine whether these differences are detectable, we need to compare them with the frequency resolution of the measurements of oscillation mode frequencies. At the edge of the frequency interval where oscillations are detected, typical uncertainties reach several tens of nHz. We thus considered here that a deviation from a regular period spacing can be detected if $\delta\nu_{\nmin}$ exceeds a threshold $\delta\nu_{\rm th} = 100$~nHz. We thus obtained the following expression for the minimal detectable magnetic perturbation
\begin{equation}
    \delta\nu_{0,\rm min} = \delta\nu_{\rm th} \left(\frac{\alpha\nmin+\beta}{ P_{\nmin}^2}+P_{\nmin}^3\numax^3\right)^{-1}
\end{equation}
We then used Eq. \ref{eq_br2min} to translate the minimal magnetic frequency shifts into minimal detectable field intensities $B_{\rm th}$, which are shown in Fig. \ref{fig_calc_B}.

\section{Can detected fields result from dynamo action in previous convective cores? \label{app_B_MS}}

The stars in which we detected strong core magnetic fields in this study have masses ranging from 1.11 to 1.56~$M_\odot$ (see Table \ref{tab_curved_RGB}). The lowest-mass stars of the sample have a radiative core during the bulk of their main-sequence evolution, but even these stars possessed a small convective core at the beginning of the main sequence because of the burning of $^3$He and $^{12}$C outside of equilibrium. We tried to determine to what extent the detected fields are compatible with dynamo-generated fields in the main-sequence convective core.

For this purpose, we assumed that a uniform field $B_{r,{\rm MS}}$ was produced during the main sequence, over a distance corresponding to the maximal extent of the convective core. After the withdrawal of the convective core, we assumed a conservation of the magnetic flux in each layer, so that the field intensity varies as $1/r(m)^2$ for a layer at Lagrangian coordinate $m$. We could then calculate the average field strength $\langle B_r^2\rangle^{0.5} = \left(\int_{\ri}^{\ro} K(r) \overline{B_r^2} \,\hbox{d}r \right)^{0.5}$ at each step of the evolution. For each star of the sample, we calculated the intensity of the main-sequence radial field $B_{r,{\rm MS}}$ that is required to produce the minimal average fields $\langle B_r^2\rangle_{\rm min}^{0.5}$ that we measured in Sect. \ref{sect_mag_strength}. We thus found main-sequence field strengths ranging from about 1 to 26~kG (see Table \ref{tab_B_MS}). These values correspond to lower limits on $B_{r,{\rm MS}}$, the actual value depending on the geometry of the current radial magnetic field. For instance, if $B_r^2$ has an axisymmetric dipolar configuration (asymmetry parameter $a=2/5$), the recovered values of $B_{r,{\rm MS}}$ range from 1 to 40~kG.

\begin{table}
\begin{center}
\caption{Minimal main-sequence field intensities required to account for current measured fields in the cores of our sample of red giants, assuming conservation of the magnetic flux along the evolution. \label{tab_B_MS}}
\begin{tabular}{ l  c }
\hline
\hline
\T \B KIC Id & $B_{r,{\rm MS}}$ (kG) \\
\hline
\T  6182668 & 18.1 \\
9474201 & 16.2 \\
6842204 & 18.7 \\
8560280 & 25.8 \\
8689270 & 19.6 \\
3216736 & 18.9 \\
5180345 & 17.8 \\
6975038 & 17.7 \\
3109742 & 5.3 \\
6614684 & 0.8 \\
\B  7728945 & 0.8 \\
\hline
\end{tabular}
\end{center}
\end{table}

\end{appendix}

\end{document}